\documentclass{elsarticle}

\usepackage{amsmath}
\usepackage{amssymb}
\usepackage{graphicx}
\usepackage{multirow}
\usepackage{multicol}

\DeclareGraphicsExtensions{.png,.pdf,.jpg} 

\usepackage{textcomp} 

\usepackage[
pdfauthor={Olivier Bourrion},
pdftitle={UCTM2: An updated User friendly Configurable Trigger, scaler and delay Module for nuclear and particle physics},
pdfsubject={UCTM},
pdfkeywords={},
colorlinks={true},
linkcolor=red,
citecolor=green,
urlcolor=cyan,
pdfstartview={FitH}
]{hyperref}

\begin{document}
\begin{frontmatter}
 
\title{UCTM2: An updated User friendly Configurable Trigger, scaler and delay Module for nuclear and particle physics}
\author[LPSC]{O.~Bourrion\corref{cor1}}
\ead{olivier.bourrion@lpsc.in2p3.fr}
\author[LPSC]{B.~Boyer}
\author[LPSC]{L.~Derome}
\author[LPSC]{G.~Pignol}

\cortext[cor1]{Corresponding author}
\address[LPSC]{LPSC, Universit\'e Grenoble-Alpes, CNRS/IN2P3 \\
53, rue des Martyrs, Grenoble, France}

\begin{abstract}
We developed a highly integrated and versatile electronic module to equip small nuclear physics experiments and lab teaching classes: the User friendly Configurable Trigger, scaler and delay Module for nuclear and particle physics (UCTM).
It is configurable through a Graphical User Interface (GUI) and provides a large number of possible trigger conditions without any Hardware Description Language (HDL) required knowledge.
This new version significantly enhances the previous capabilities by providing two additional features: signal digitization and time measurements.
The design, performances and a typical application are presented.

\end{abstract}


\end{frontmatter}
\section{Introduction}
Earlier, a multi-purpose ``User friendly Configurable Trigger, scaler and delay Module'' (UCTM1)\cite{uctm1} providing configurable logic functions and delay capability for trigger building as well as scalers was designed to equip small nuclear experiments or for teaching in lab classes.
The module was composed of two interdependent parts: a configurable electronics board relying on a FPGA for the digital functions and a Graphical User Interface (GUI) to be usable by physicists or students having no particular knowledge in any Hardware Description Language (HDL).
A new module (UCTM2), keeping all triggering\footnote{At each run start up, the equations are used to compute each output truth table which are loaded in the FPGA memory blocks. Note that consequently the FPGA firmware is never modified, so there is no need for synthesis and  placer-router tools to configure the board, thus avoiding any tool licensing issue.} and counting features from the first design, was designed to provide new features and to have improved performances.

\begin{figure}[ht]
\begin{center}
\includegraphics[angle=90,height=7cm]{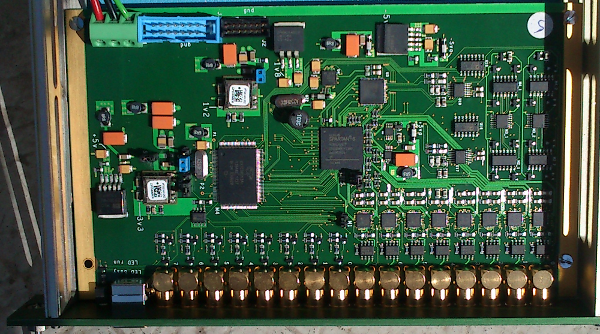}
\hspace{1cm}
\includegraphics[angle=90,height=7cm]{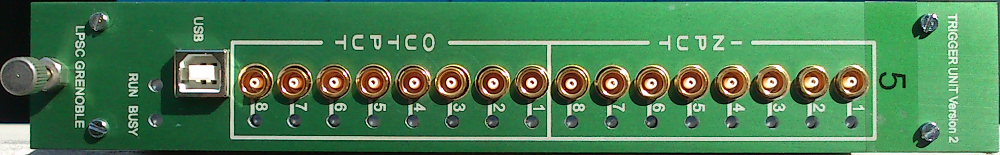}
\caption{Picture of the electronic board inserted in the module (left hand
side) and of the front panel (right hand side).}
\label{photocarte}
\end{center}
\end{figure}
The UCTM2, shown in fig.~\ref{photocarte}, fits in a NIM module and features eight analog inputs and eight configurable digital outputs. 
It can perform up to eight simultaneous time measurements between any logical combinations of the inputs with a resolution of 5\,ns.
Additionally, two out of the eight inputs can be selected by the user to be digitized at 200\,Msps by 12 bit Analog to Digital Converter (ADC).
The digitized signals can then be directly read-out and/or processed to provide peak amplitude or charge measurement. The trigger or the gating signal used for these operations being again any logical combinations of the inputs.

The GUI associated the UCTM2 electronics is used to set-up the trigger equations and to read-out the scalers.
It also provides visualization tools such as ``multi-channel analyzer windows'', a ``Time to digital converter histogram window'' and ``oscilloscope windows''.

This paper is organized as follows: section~\ref{HardwareSec} presents the hardware design, section~\ref{FPGASec} describes the FPGA contents. The control and readout software is presented in section~\ref{SoftSec}.
Eventually, a typical example application is described in section~\ref{ExampleApplications} and a short summary is given in section~\ref{SummarySec}.
\section{Hardware description}
\label{HardwareSec}

\begin{figure}[ht]
\begin{center}
\includegraphics[angle=0,width=0.9\textwidth]{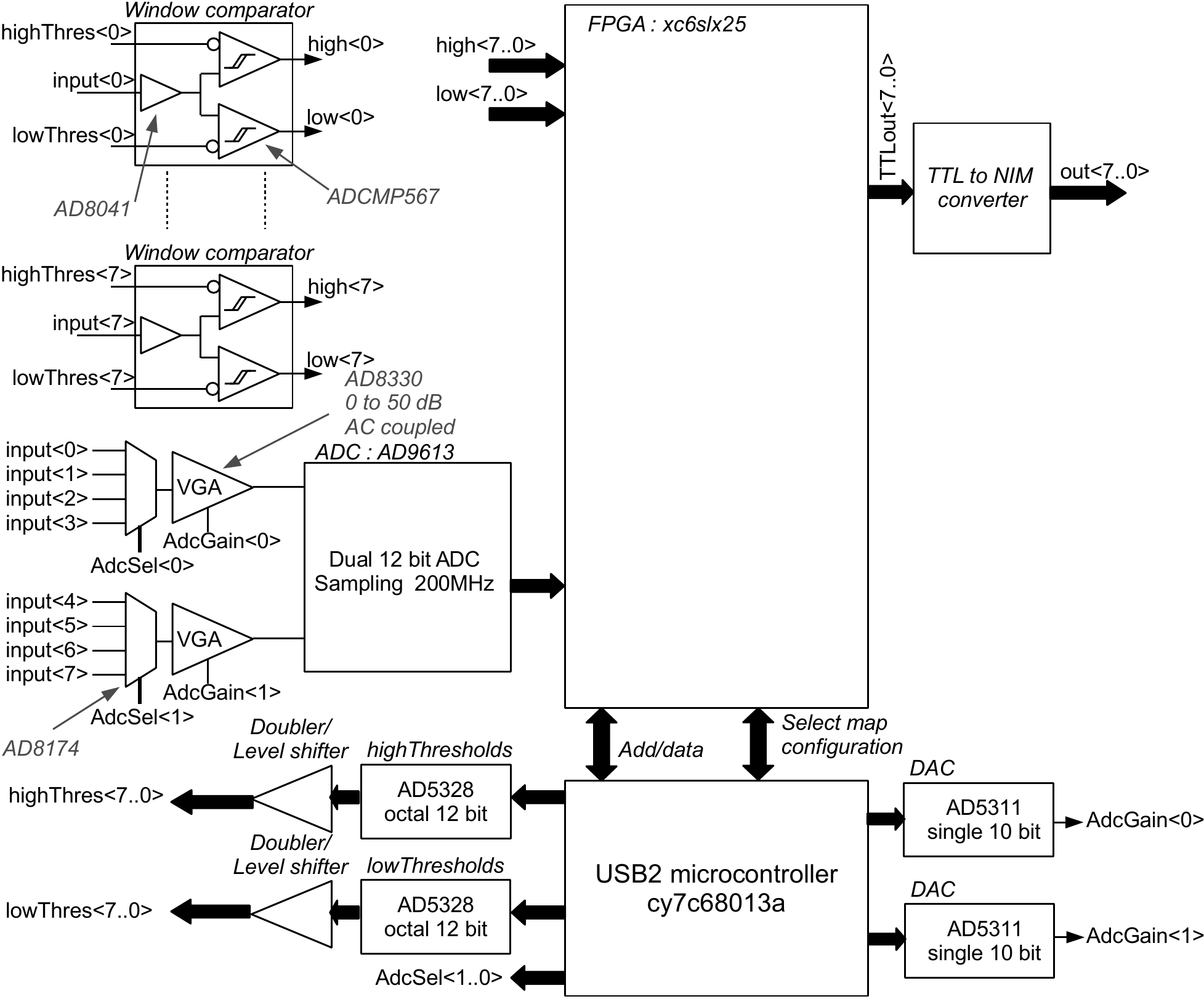}
\caption{Electronics overview.}
\label{HardBlockDiag}
\end{center}
\end{figure}
As shown in fig.~\ref{HardBlockDiag}, the system is built around a FPGA (XC6SLX25-2CSG324), hosting the digital functions, and a micro-controller (CY7C68013A) used to provide a Universal Serial Bus (USB) interface with the control and readout computer.
The micro-controller drives the various DAC used for gain and threshold settings, the ADC input multiplexers.
It is also used to configure the FPGA internal registers and to provide the FPGA firmware at system start-up.

On the input side, each analog signal is connected to a window comparator and to a multiplexer.
The window comparators, that provide the discriminated inputs to the FPGA, are made of two fast comparators (ADCMP567) and of a buffer used to duplicate the input signal.
Each window comparator threshold (low and high) can be adjusted between -3.3\,V and +3.3\,V with a resolution of 1.61\,mV, thank to the 12 bit Digital to Analog Converters (DAC).

Two multiplexers (AD8174) are used to collect the eight analog inputs and to select the two channels to be digitized after amplification by Variable Gain Amplifiers (VGA) (AD8330).
The VGA gains can be independently adjusted between 0\,dB and 50\,dB by two 10 bit Digital to Analog Converters (DAC) (AD5311).
The VGA inputs are AC coupled in order to remove any offset that may otherwise saturate the VGA amplifiers in case of large gain.
The AC coupling has a low frequency cut-off of $\rm f_{-3\,dB}=32\,$Hz.

The dual ADC (AD9613) operated at 200\,Msps has a 12 bit resolution and an input dynamic range of 1.75\,Vpp.

To summarize the input specifications, the discriminator path is DC coupled and has an input range of -3.3\,V to +3.3\,V, while the digitization path is AC coupled and has a maximum input range of -0.875\,V to +0.875\,V, that is when VGA gain is set at 0\,dB.

On the output side, the FPGA digital signals are converted to the NIM standard with the system described in \cite{uctm1}.

\section{Firmware description}
\label{FPGASec}
\begin{figure}[ht]
\begin{center}
\includegraphics[angle=0,width=0.99\textwidth]{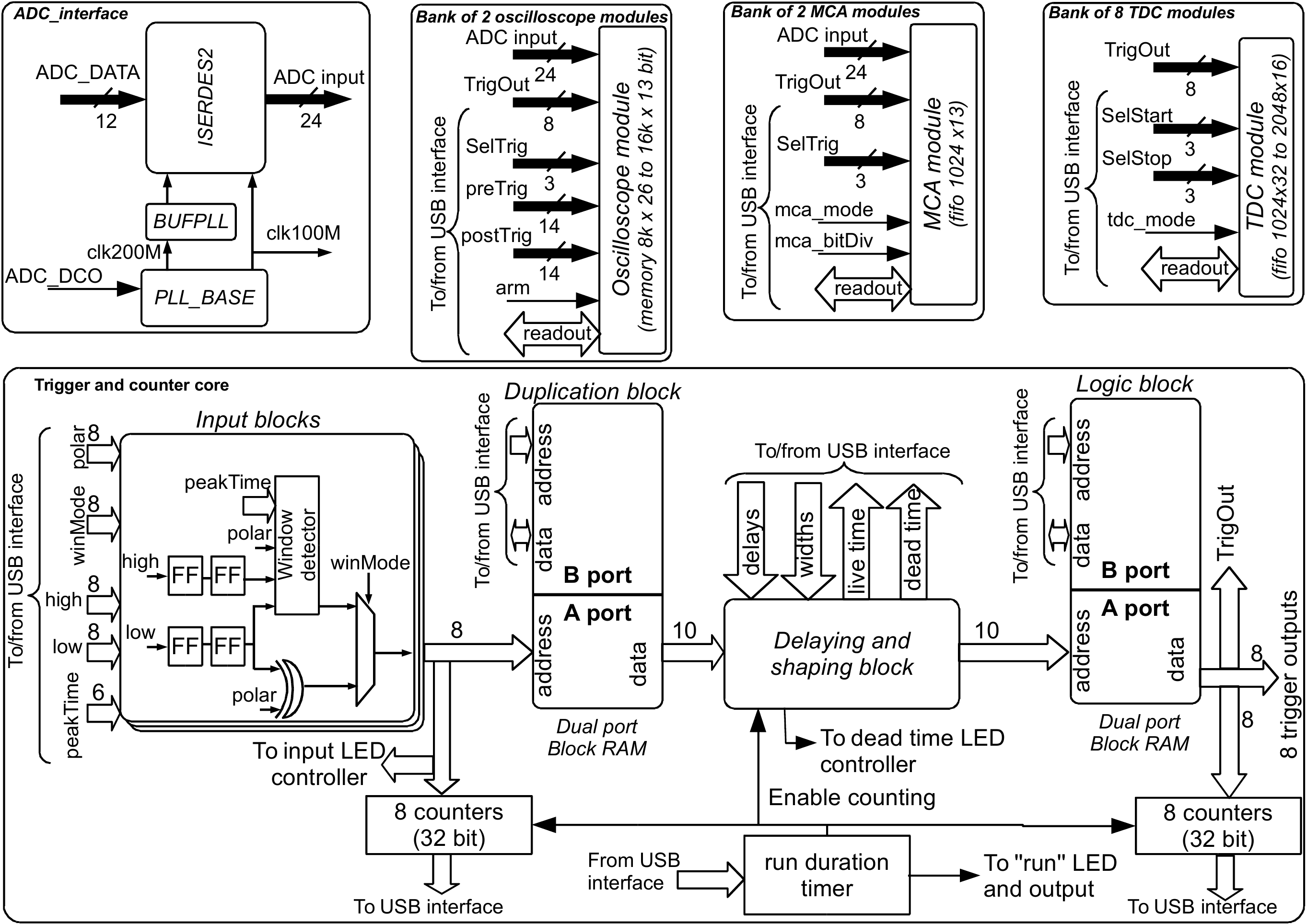}
\caption{Firmware overview. It is composed of five parts:  the trigger and counter core, the ADC interface, two multi-channel analyser (MCA) modules, eight time to digital converter (TDC) modules and two oscilloscope modules. The LED controllers and the USB micro-controller interface used to configure the blocks are not shown}
\label{FirmBlockDiag}
\end{center}
\end{figure}

An overview of the FPGA firmware is given in fig.~\ref{FirmBlockDiag}.
It is composed of five parts:  the trigger and counter core, the ADC interface, two multi-channel analyser (MCA) modules, eight time to digital converter (TDC) modules and two oscilloscope modules. The LED controllers and the USB micro-controller interface used to configure the blocks are not shown. 

\subsection{Trigger and counter core}
The ``trigger and counter'' core is composed of the input blocks, the duplication block, the delay/shaping block and the logic block. 
The trigger path, from input to output is clocked at 200\,MHz.

The 16 signals provided by the window comparators are first fed to the input blocks.
Each of these, synchronizes the signals with two input flip-flops (FF) before directing them to the polarity inverter (a XOR gate) for a direct use and to the digital window detector.
The window detector analyses the signal and provides a trigger output only if the signal has not crossed the high level threshold before a duration of ``peakTime'' after it has crossed the low level threshold.
Consequently, in the window mode, there is an additional latency  of the programmed peak time for taking the trigger decision. 
``peakTime'' can be configured from 1 to 63 clock cycles, i.e up to 315\,ns.
Each input block comprises a multiplexer used to select the trigger source to forward to the duplication block, the 32 bit counter and the LED controller.

The ``duplication block'', which is equivalent to a fanout buffer, uses eight
inputs and can provide up to ten signals to the downstream blocks. It is used to
replicate a chosen input a preprogrammed number of times. This function is
achieved by using the concatenated input signals as an eight bit address  which
points to a ten bit word containing the duplication result. This $2^8 \times
10$ bits duplication matrix is computed beforehand for all input combinations to
eventually associate each address bit to one or several data outputs.

The ``delay/shaping'' block is used to delay the signal from 0 to
\mbox{2\textsuperscript{16}-1} clock cycles and to adjust its width within a
range of 1 to \mbox{2\textsuperscript{16}-1} clock cycles.
This function is performed by two adjustable monostables (one for delaying and one for width adjustment). Provided that the input signal edge is used as the reference time to perform the delaying and shaping, a signal width can be reduced as well as enlarged.
Any new input pulse received by a delaying/shaping block while the previous processing is still in progress is ignored.
Consequently, the larger the delay/width programmer, the longer the dead time.
To monitor this effect, 32 bit dead time and live time counters are implemented for each shaping channel.

The last and most important block of the firmware is the ``logic block''
which behaves in a similar way as a Look Up Table (LUT) in any FPGA.
Similarly to the ``duplication block'', the ``logic block'' is preloaded with a truth table giving the expected output vector as a function of the input vector.
The input vector, composed of the ``delay/shaping block'' output is used for the memory address and the memory data is used as the output vector.
Hence, each memory output bit is a function of up to ten inputs (or operands)
and the latency of this block is independent of the trigger equation
complexity.
The memory block data output directly reflects the expected behavior when using the two input bit as the address vector and by conveniently preloading the memory content with the OR gate truth table.

The eight trigger outputs are used internally for the MCA, TDC and oscilloscope modules and provided for external use.
Additionally, the trigger output rates are monitored by 32 bit counters.

\subsection{ADC interface}
The ADC uses a 6 bit dual data rate (DDR) differential interface to transfer the samples at 200\,MHz to the FPGA.
To cope with this rate and to ease the design and implementation of the processing modules, the data are deserialized with the ISERDES2 included in the FPGA fabric and provided as two bus of 12 bit data (ADC\_input) clocked at half of the input frequency, i.e 100\,MHz.
The ADC data clock (ADC\_DCO) is used to generate the reference de-serialization clock and the system clock, thanks to phase locked loop (PLL) and a dedicated input clock buffer (BUFPLL).

\subsection{Oscilloscope module}
Each oscilloscope module can record up to 16384 data samples, thanks to the usage of large memory blocks.
This represents 81.82\,\textmu s of recording at 200\,MHz.

The memory is implemented in a dual port memory having different data bus write and read sizes.
The input port of the memory is used to record the two bus of ADC data (two times 12 bit) provided by the ``ADC\_interface'' and the trigger used (2 bit); this recording takes place at 100\,MHz.
The output port is used by the ``USB interface'' to read the 12 bit ADC data concatenated with the trigger bit.

The number of samples to record before (preTrig) and after the trigger (postTrig) can be adjusted independently for each module, as long as the total sample count does not exceed 16384.
The trigger used is selected  by the ``selTrig'' port among those generated by the ``trigger and counter core''.

\begin{figure}[ht]
\begin{center}
\includegraphics[angle=0,width=0.4\textwidth]{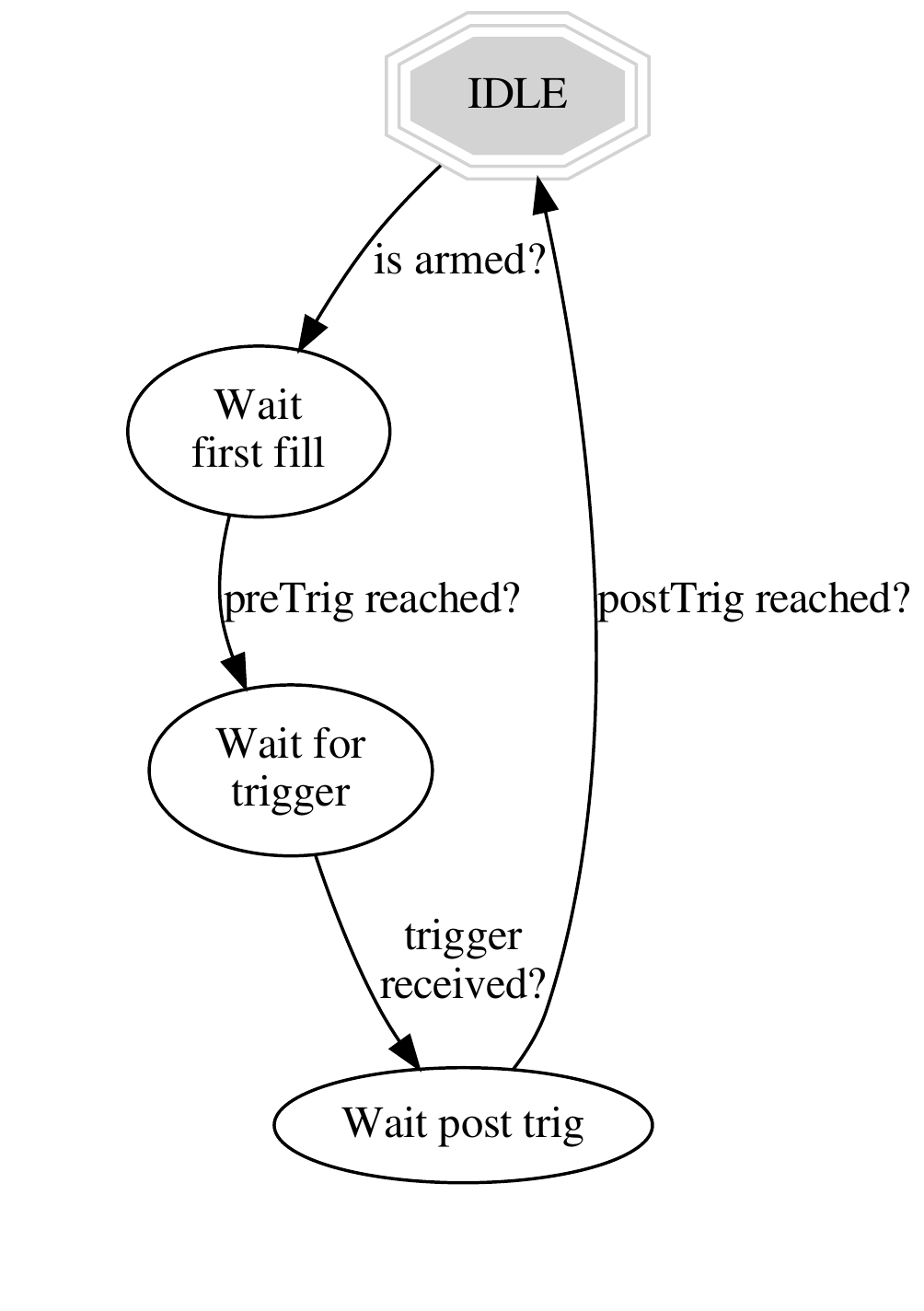}
\caption{Oscilloscope module recording finite state machine.}
\label{oscFSM}
\end{center}
\end{figure}
The data recording is controlled by a finite state machine (FSM), shown in figure~\ref{oscFSM}.
Once armed via the ``USB interface'', the ADC data are continuously written in memory.
Then, after ensuring that the required ``preTrig'' data are recorded in the ``wait first fill'' state, any trigger can be accepted.
Once a trigger is received, the FSM writes ``postTrig'' more data in the memory before disabling the write port and then returns to the ``IDLE'' state.

It must be noted that the trigger generated at 200\,MHz is stretched to be detected by the FSM operated at 100\,MHz.
The fact that an uncertainty of two samples (or 10\,ns) is introduced is considered negligible, provided that the recorded data can be used to determine off-line the actual triggering sample.

\subsection{Multi-channel analyser}
The MCA uses the samples provided by the ``ADC\_interface'' to either perform minimum, maximum, amplitude measurements or digital integration during a time controlled by a gate signal.
This gate signal is selected among the eight ``trigOut'' signals provided by the ``trigger and counter core''.
The type of measurement is selected by the mca\_mode bit.

The gate signal adjustment, done with the ``delaying and shaping block'', can be tricky.
To ease this tuning, the ``oscilloscope module'' should therefore be used during the set-up time to visualize the relative timing of the gate signal with respect to the input signal.

MCA operation is controlled by a FSM, that performs the operations according to the selected ``mca\_mode'' and writes the data generated in a memory buffer having a size of 1024 words of 16 bit.
To ease the driving software implementation, the FSM always provides the absolute value of the resulting measurement on 13 bit.
The additional bit, with respect to the ADC bus resolution (12 bit), is required to avoid eventual overflow in the amplitude measurement mode.
Eventually, the FIFO is read out by the ``USB interface''.

In the integration mode, an internal integrator having a width of 21 bit is used.
Hence, to avoid digital clipping, some LSB must be dropped on the total integrated value to fit ``MCA module'' output width of 13 bit.
This parameter, controlled by the ``USB interface'', is provided by the ``mca\_bitDiv'' port.

\subsection{Time to digital converter}
The TDC module features two operating modes.
In the ``single stop'' a time is measured between a start signal and a stop signal, i.e. for each start a single measurement is made.
In the ``multi-stop'' all times from a single start to all subsequent stop signals are measured, i.e. for each start an undetermined number of data can be produced.

As for the previous modules, any of the trigger signal generated by the ``trigger and counter core'' can be used as start and stop signal. 
The start and stop selection as well as the TDC operation mode are done via the ``USB interface''.
Each TDC mode is configured independently.

TDC operation is controlled by a FSM.
Upon detection of a start signal, an internal 24 bit counter is cleared and allowed to be incremented by the 200\,MHz clock.
At the reception of a stop signal, the counter value is written in the output FIFO. 
Then, if in ``single stop'', the counter is stopped and any further stop signal is ignored. 
Alternatively, when in ``multi stop'', the operation continues until a new start signal is detected and for each subsequent stop signal a new counter value is written in the output FIFO.
Provided the clock frequency and the counter size, the dynamic range of the time measurement is from 5\,ns to about 83\,ms.

\subsection{Resource usage}
The firmware fits in XC6SLX25-2CSG324 FPGA.
It uses 88\% of the memory blocks, 30\% of the slice registers and 46\% of the slice look-up tables for a total of 69\% of occupied slices. 
The memory block usage is summarized in table~\ref{memTable}.

\begin{table}
\begin{center}
\caption{Memory resources used in the XC6SLX25-2CSG324 FPGA.\label{memTable}}
\begin{tabular}{|c|c|}
\hline
  Module & memory blocks used (RAMB16WER) \\
\hline
  2 oscilloscopes & $2 \times 13 = 26$\\  
\hline
 2 MCA & $2 \times 1 = 2$\\  
\hline
  Duplication block & 1\\  
\hline
  Equation block & 1\\  
\hline
  8 TDC & $8 \times 2=16$\\  
\hline
\hline
 Total & 46 \\
 \hline
\end{tabular}
\end{center}
\end{table}

\section{Readout and control software}
\label{SoftSec}
The readout and control software, which is written in C++, is composed of two layers: the Application Programming Interface (API) and the application software.

The API uses open source software drivers to provide hardware control over the USB port \cite{libusb}.
Aside from providing the basic functionality for accessing the electronics memory map, it also provides the trigger equation parser, the truth table and duplication matrix generators.
The main building block of the API is the equation interpreter and computer.
It is designed to interpret equations with up to ten operands and to manage all the basic binary operators, such as AND, OR, XOR (exclusive OR), XNOR (complemented exclusive OR), NOR and NAND, as well as the unary operators: NOT() and SUP().
SUP() is the multiplicity operator, accepting a list of operands and the target
multiplicity number.
A full description of the equation interpreter implementation as well as trigger equations examples can be found in \cite{uctm1}.
\\

The Graphical User Interface (GUI) described hereafter is an example implementation using all the capabilities of the electronics. 
A customized application software using only the required features for a lab set-up can be written.
The example GUI software \cite{Qt,Qwt} is implemented as a Multiple Document Interface (MDI), where a child window is shown for each module activated.
To ease the electronics usage, efforts were made to design an intuitive GUI. 
A readout loop is executed every 10\,ms and all modules activated are included in this loop, hence, the maximum readout rate is configuration dependent.

\begin{figure}
\begin{center}
\includegraphics[angle=0,width=0.99\textwidth]{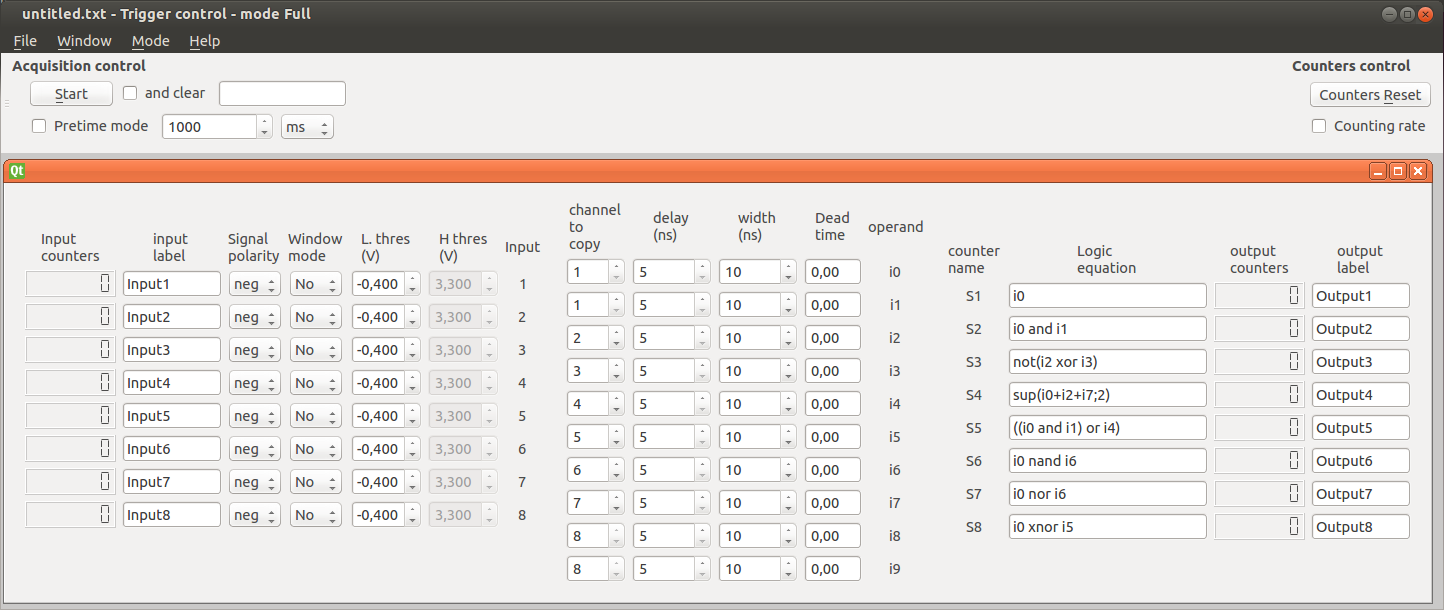}
\caption{Screenshot of the main window containing the mandatory child window controlling the ``trigger and counter core''.\label{mainPic}}
\end{center}
\end{figure}

The parent window which provides the acquisition and the counter controls and contains the child window controlling the ``trigger and counter core'', which is a mandatory module, is shown in fig.~\ref{mainPic}.
The child window is divided in three zones. 
On its l.h.s appears the input scalers, the discriminator mode selector, thresholds tuning and their user selectable labels.
The middle part is used to select the duplication/routing and to enter the delaying/shaping settings.
Percentage of dead-time is provided for duplicated and shaped channel.
Finally, on its r.h.s eight trigger equations, using one up to ten operands, can be entered with their associated labels and activity counters.

\begin{figure}[ht]
\begin{center}
\includegraphics[angle=0,width=0.8\textwidth]{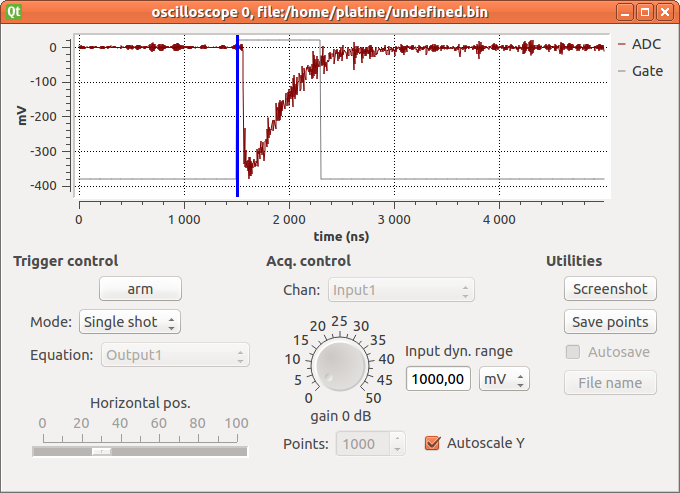}
\caption{Screenshot of an ``oscilloscope'' child  window. The vertical blue line represents the trigger position, the trigger signal is shown in grey. It can be used as a gate for the MCA and in red the recorded ADC signal is shown.\label{oscPic}}
\end{center}
\end{figure}

An ``oscilloscope'' child  window  screen capture can be seen in fig.~\ref{oscPic}. 
The upper part is used to display the captured waveform as well as the gate/trigger signal associated.
The cursor line is used to represent the trigger position.
The lower part is composed of three parts.
On its l.h.s appears the trigger control: mode selection (single shot or normal), trigger source (selectable by label name) and the trigger horizontal position.
In the middle, the user can select by label the channel to the digitize, the gain (via the knob or by entering a value) and the number of points to acquire.
On its r.h.s, some utilities are provided to capture a waveform in the form of image or data points.
When selecting ``autosave'', all consecutive waveform data points are stored in a single file for later analysis.

\begin{figure}[ht]
\begin{center}
\includegraphics[angle=0,width=0.8\textwidth]{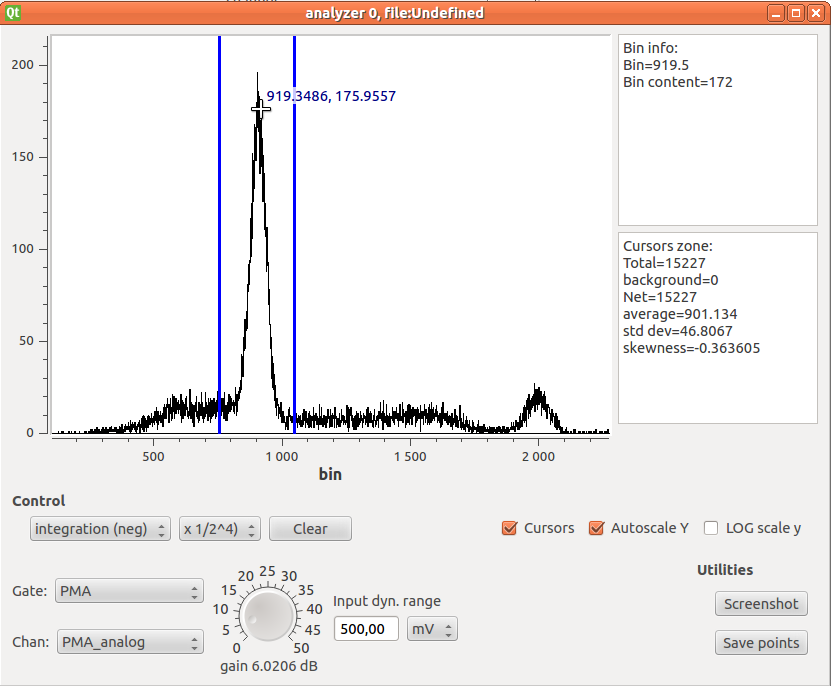}
\caption{Screenshot of an ``MCA'' child  window. 
In this example, the gamma rays emitted by a $^{22}$Na radioactive source are detected with a scintillator coupled to a photomultiplier. 
The amplitude spectrum of the PMT pulses are analyzed with the UCTM2 unit. 
\label{mcaPic}}
\end{center}
\end{figure}

\begin{figure}[ht]
\begin{center}
\includegraphics[angle=0,width=0.9\textwidth]{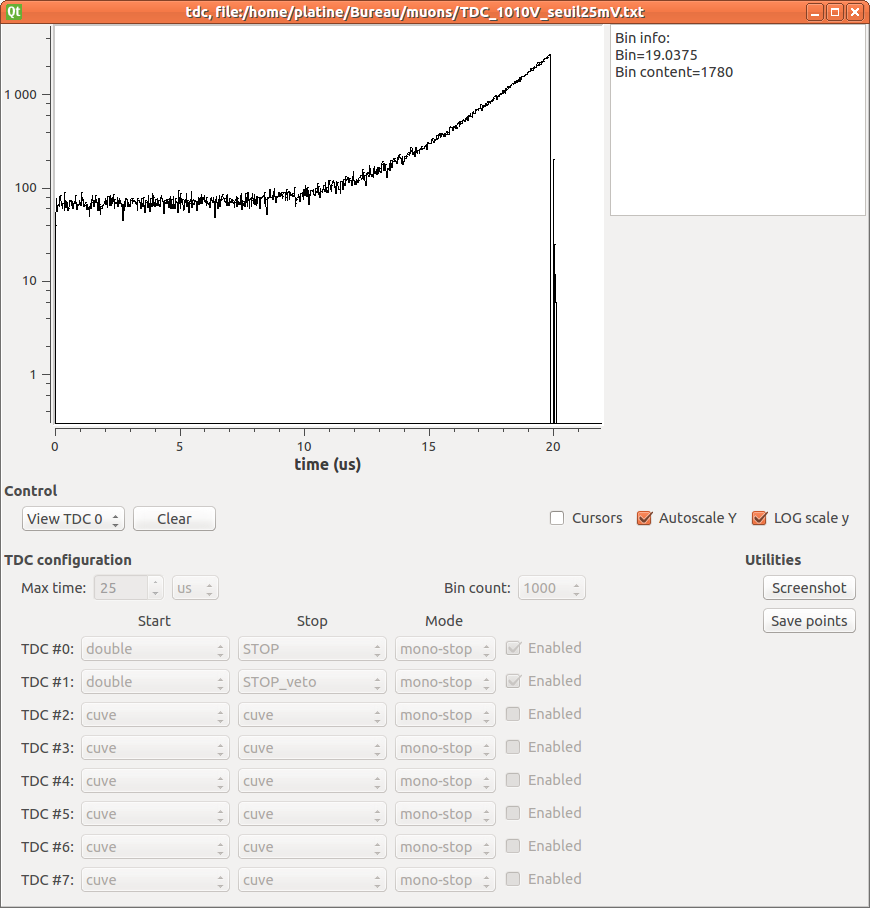}
\caption{Screenshot of a ``TDC'' child  window.
In this example, the muon lifetime is measured with a procedure described in section~\ref{ExampleApplications}.
\label{tdcPic}}
\end{center}
\end{figure}

Screenshots of an MCA and of a TDC child window can be seen in fig.~\ref{mcaPic} and fig.~\ref{tdcPic} respectively.
Both of them have an histogram on their upper parts to display the data recorded.
A light analysis tool is implemented to provide some information on the histogram part located between two movable cursors.
The control of the MCA child window, located on the lower part, allow to choose the histogram type  (positive amplitude, negative amplitude, total amplitude or charge),  the channel input, the gating signal and the gain applied.
Note that the channel selected and its associated gain is shared with the corresponding oscilloscope window.
The control of the TDC child window, also located in the lower part, allows to choose which TDC information to display, the bining and dynamic range.
Then for each enabled TDC channel, the start/stop signals as well as the operating mode can be chosen.

\section{Example application: muon lifetime}
\label{ExampleApplications}

\begin{figure}[ht]
\begin{center}
\includegraphics[angle=0,width=0.99\textwidth]{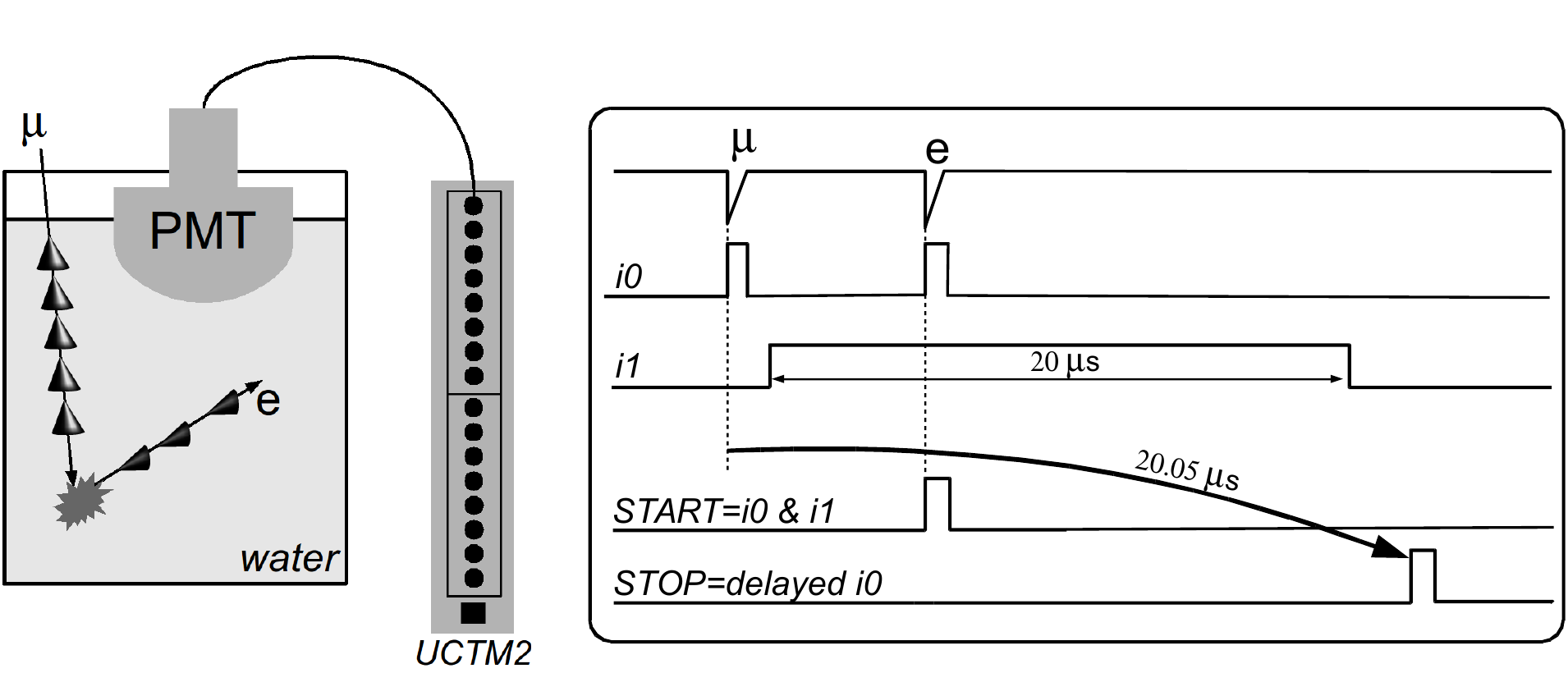}
\caption{Sketch of the muon lifetime measurement set-up, see text for details. 
\label{experimentSketch}}
\end{center}
\end{figure}

As an illustration of the capabilities of the UCTM2 unit, we describe a simple lab experiment designed to measure the muon lifetime \cite{Hall1970}. 
The experiment uses the flux of muons from secondary cosmic rays, detected by their Cerenkov light produced in a water tank ($40\times40\times60$\,~cm\textsuperscript{3}). 
The light is recorded with a photomultiplier tube (PMT) immersed in the water tank (see fig.~\ref{experimentSketch}). 
Cosmic muons with an incident kinetic energy in the range $50 \, {\rm MeV} \lesssim E \lesssim 150 \, {\rm MeV}$ are fast enough to produce Cerenkov light and slow enough to be brought at rest in the tank. 
After an average time of $\tau_\mu \approx $ 2.2\, \textmu s, a muon decays into an electron (positive or negative) and a pair of neutrinos: $\mu \rightarrow e + \bar{\nu} + \nu$. 
Most of the time the electron velocity exceeds the Cerenkov threshold and therefore is also seen by the PMT. 

The measurement consists in analyzing the time difference between the ``muon'' and ``electron'' pulses in the PMT signal, following the standard methodology: 
\begin{itemize}
\item The analog PMT signal is discriminated with a low edge discriminator, producing a short logic gate (i0) for each pulse. 
\item A 20\,\textmu s long secondary gate (i1) triggered by i0 is produced. 
\item The coincidence between i0 and i1 forms the START signal. 
\item The STOP signal is formed by delaying the i0 signal. 
\item The START and STOP signals are used as inputs for the internal TDC module. 
\end{itemize}
This procedure is sketched in fig.~\ref{experimentSketch}. 
Implementing the above scheme would require several nuclear instrument modules: 
a discriminator, two delay modules, a coincidence unit, a TDC and a multichannel analyzer. 
Here we used the UCTM2 unit to replace all electronic modules. 
Figure~\ref{tdcPic} shows the TDC output after eight days of data taking, the detection rate is about 0.5\,Hz.
The histogram consists of a constant background and of an exponential decay with a typical time of 2\,\textmu s.
The constant background corresponds to the accidental detection of two crossing muons in the 20\,\textmu s time window.
The exponential decay is a clear signature of the muon decay.

\section{Summary}
\label{SummarySec}
A configurable trigger scaler and delay NIM module has been designed to equip nuclear physics experiments and lab teaching class.
It is configurable through a Graphical User Interface (GUI) and provides a large number of possible trigger conditions without any Hardware Description Language (HDL) knowledge.

The module has eight inputs that can be discriminated and two of these, user selected, can be digitized at 200\,Msps by 12 bit Analog to Digital Converter (ADC).
The input discriminators have individually configurable low and high thresholds and can operate in a windowed mode.
The discriminated version of the inputs can be logically duplicated and used in trigger equations that are entered as plain string in the control and readout GUI.
Each trigger equation is provided as output in the NIM standard and can be used to control internal oscilloscope, multi-channel analyzer and time to digital converter modules.
Scalers are available on each input and output.

\begin{table}
\begin{center}
\caption{Summary of module specifications.\label{specTable}}
\begin{tabular}{|c|l|c|}
\hline
  Function & Parameter & Value \\
\hline
\hline
  \multirow{3}{*}{Analog inputs} & Total number & 8 \\
  & Usable by trigger & all \\
  & Usable by digitizer & 2 \\
\hline
  \multirow{5}{*}{Input discriminators} & Number & 8 \\
  & Dynamic range & -3\,V $\rightarrow$ +3\,V \\
  & Signal coupling & DC \\
  & Mode & simple/windowed \\
  & Tuning resolution & 1.6\,mV \\
\hline
  \multirow{3}{*}{Shaping} & Delay range & 0\,ns $\rightarrow$ 327.38\,\textmu s \\
  & Width range & 5\,ns $\rightarrow$ 327.38\,\textmu s \\
  & Tuning resolution & 5\,ns \\
\hline
  \multirow{3}{*}{Trigger equation} & count & 8 \\
  & Operand count & 10 \\
  & Binary operators & and, nand, or, nor, xor, xnor \\
  & Unitary operators & not, multiplicity \\
\hline
	Outputs & NIM standard & 8\\
\hline
  \multicolumn{2}{|c|}{Minimum input to output latency}  & 35\,ns\\
\hline
  \multirow{5}{*}{ADC} & Sampling frequency & 200\,MHz \\
  & Resolution & 12 bit \\
  & Signal coupling & AC ($\rm f_c=32\,Hz$) \\
  & Dynamic range & 1\,Vpp \\
  & Variable gain & 0\,dB $\rightarrow$ 50\,dB \\
\hline
  \multirow{5}{*}{Oscilloscopes} & Number & 2 \\
  & Sampling frequency & 200\,MHz \\
  & Trigger accuracy & 10\,ns \\
  & Memory depth & 16384 points \\
  & Mode & simple/windowed \\
\hline
  \multirow{5}{*}{MCA} & Number & 2 \\
  & Buffer size & 1024 points \\
  & Resolution (bin count) & 12 bit (4096) \\
  & Modes & amplitude, min., max. \\
  &       & integration \\
\hline
  \multirow{5}{*}{TDC} & Number & 8 \\
  & Buffer size & 1024 points \\
  & Stop mode & single/multiple \\
  & Resolution & 5\,ns \\
  & Dynamic range & 5\,ns $\rightarrow$ 83\,ms \\
\hline
\end{tabular}
\end{center}
\end{table}

\end{document}